\documentclass[fleqn,twoside]{article}
\usepackage{espcrc2}
\usepackage{graphicx}
\newcommand{\AmS}{{\protect\the\textfont2
  A\kern-.1667em\lower.5ex\hbox{M}\kern-.125emS}}

\title{Extensive Air Showers and the Physics of High Energy Interactions}
\author{A.D.Erlykin\address{P.N.Lebedev Physical Institute,   
        Leninsky prosp. 53, Moscow 119991, Russia}
        \thanks{E-mail: erlykin@sci.lebedev.ru}}

\begin{document}

\begin{abstract}
Extensive Air Showers are still the only source of information on primary cosmic rays 
and their interactions at energies above PeV. However, this information is hidden 
inside the multiplicative character of the cascading process. Inspite of the great 
experimental and theoretical efforts the results of different studies are often 
ambiguous and even conflicting. These controversies can partly be referred to 
imperfections of our models of high energy particle interactions. 

The first part of the paper is concerned with this problem. The author thinks 
that the present models should be corrected to give slightly deeper penetration of the 
cascade into the atmosphere. In this respect the modification suggested by the QGSJET-
II model seems to be the step in the right direction. The Sibyll 2.1 model provides a 
similar penetrating properies. However, this modification is not
enough and a small additional transfer of the energy from EAS hadrons
to the electromagnetic component is needed too. As a possible
candidate for such a process the inelastic charge exchange of pions is discussed. 

In the second part of the paper the author discusses the need to account for the 
interaction of EAS  with the stuff of detectors, their environment and the ground in 
the light of the 'neutron thunder' phenomenon, discovered recently.  
\end{abstract}

\maketitle

\section{Introduction}

There is a big progress in the analysis of experimental data
on extensive air showers (~EAS~) during the last two decades. However, 
one cannot say that we understand all the phenomena and
characteristics of EAS which we
observe. Partly this dissatisfaction is due to the controversies in
experimental data themselves, partly due to still remaining
imperfections of the analysis. We certainly need to improve our understanding 
of EAS. 

This paper does not aim to give a comprehensive review of all high
energy interaction models, event generators and EAS simulation codes.
It consists of two different parts. In the first part I point out some problems 
related to the particle interaction models which so far pose questions 
at high energies. I do not go into the theoretical foundations  
of various interaction models, but stay within a pure
phenomenological approach. Within it I indicate the possible way
to improve the models. The theoretical basis of some recent models can
be found in \cite{Stane}. 

In the second part of the paper I shall touch the problems related to some 
effects of the EAS interaction with the environment. 

\section{High energy interactions}

\subsection{The consistency of the results} 
The EAS is a complex phenomenon - it has several different components: electromagnetic - 
electrons, positrons and gamma-quanta, muons and hadrons - nucleons, pions, kaons and so on. 
Besides that there are neutrinos which need massive detectors to be studied. Due to their 
small interaction cross-section they are detected not as multiple shower neutrinos, 
but as single ones. So far they are {\em not} combined with other EAS components in the 
analysis of experimental data, but they certainly play a role in the
energy balance. Optical cherenkov and fluorescence photons emitted by
charged shower particles are also used as a powerful tool for the study.  

Since the characteristics of observed showers are the product of the
primary cosmic-ray (~CR~) energy 
spectrum, mass composition and high energy interactions, the only way to disentangle 
them is to achieve the self-consistency in the derivation of the properties of primary 
CR from different observables and vice versa - the derivation of observed 
characteristics for different shower components and different observation levels from
the same primary CR and the interaction model.  
     
There were many efforts in the past to use models of the popular CKP- or scaling type.
With the development of the QGS model \cite{Kaida} it has been shown that this model gives 
a satisfactory description of both EAS \cite{Kalm1,Kalm2} and single, unassociated CR components 
in the atmosphere \cite{Erly1,Erly2,Erly3}. 
However, those old studies used as a rule different cascading algorithms and 
programs, which certainly produced an additional uncertainty in the results and reduced
 the credibility of the conclusions. It is to the credit of the KASCADE people 
who spend great efforts to develop and to distribute freely the CORSIKA code \cite{Heck}. 
With this code the analysis of experimental data can now be made at the level much better 
than before. 

\subsection{The improvement of models}

An early analysis of models indicated that the best consistency for the mean 
logarithmic mass $\langle lnA \rangle$ of primary CR derived from the $\frac{N_\mu}{N_e}$ 
ratio and from $X_{max}$ can be achieved for the QGSJET model \cite{EW1}. Here $N_\mu, N_e$
are muon and electron sizes of EAS respectively and $X_{max}$ - the depth of the shower
 maximum. Later a similar conclusion about the preference of QGSJET model has been 
made on the basis of the analysis of the EAS hadronic core \cite{Anton}. After some 
improvements the SIBYLL model, version 2.1 joined the list of the best, most popular
 and often used models \cite{Engel}. 

However, the closer look reveals that some inconsistencies still
remain. It has to be said 
that indications of possible inconsistencies appeared more that 30 years ago when
the mismatch between the direct and indirect measurements of the primary energy 
spectrum has been noticed: the indirect measurements based on the EAS model 
calculations gave as a rule the higher CR intensity in the PeV region than that derived
 by the extrapolation of direct measurements from the lower energies - the so called 
'bump' problem \cite{Wdowc}. More recently this mismatch has been confirmed by
\cite{Hoer1}. Among possible explanations there was an assumption that even the best 
models give an overestimation of the primary energy from the observations in the 
atmosphere. It could happen if the shower penetrates deeper into the atmosphere
and has more charged particles at the observation level than it is expected from model 
calculations. 

    Observations of the EAS cherenkov light in the PeV region confirmed this deeper 
penetration \cite{Hoer2}. As a consequence, the primary mass attributed to such showers
derived from observed $X_{max}$ values and $N_\mu/N_e$ ratio after the comparison 
with model calculations turned to be smaller than the true primary mass. There was a 
number of ideas how to increase the penetrability of the showers, for instance, 
introducing the higher cross-section for the charm production \cite{Yakov} or  
hypothetical strangelets \cite{Rybcz,Shaul}, but those models are still in the stage of 
development. The possibility to improve the models were discussed also in 
\cite{Hoer2,EW2}. In \cite{Hoer2} it has been assumed that the cross-section and the 
inelasticity of the 
proton interactions in the air are in fact smaller than in the models, although they 
still agree with measurements at the lower end of the error bars. Their reduction 
allowed to improve the agreement between the predictions of the models and the results 
of the $X_{max}$ measurements. There were some indications of 
the lower cross-sections in the past measurements of hadrons in the EAS cores 
\cite{Neste}. The latest measurements of the inelastic cross-sections confirmed the 
slower rise of the interaction cross-section with energy \cite{Aglie,Belov}. Therefore,
 there are experimental indications that EAS may in fact penetrate deeper, than 
predicted by models.

There are also efforts to improve models not just on the pure phenomenological, but 
also on the theoretical basis. The idea that the density of partons at high energies   
becomes so high that they cannot interact independently of each other has been 
discussed long ago \cite{Aniso}. However, it is to the credit of S.S.Ostapchenko, who 
updated the QGSJET01 model including the non-linear effects of parton interactions,  
developed it to the status of the Monte Carlo event generator and together with his 
colleagues in Karlsruhe incorporated it into the Corsika code \cite{Osta1,Osta2}. 
As a consequence of the non-linear effects, the interaction cross-section (~at least 
for pions~), the multiplicity of secondaries and the inelasticity of the 
collisions decreased slightly which helped atmospheric cascades to penetrate deeper.
Apparently the reduction of the inelasticity plays the major role in the increased 
penetrability.
Due to its smaller inelasticity the updated Sibyll 2.1 model also provides EAS with 
a greater penetrability than previous models. Certainly these improvements are 
the step in the right direction.

However, the only introduction of the non-linear effects of parton 
interactions seems to be not enough. This suspicion appears when the examination of the
 hadron component is included
 into the analysis. It has been shown in \cite{Roth} that the primary mass composition 
derived mainly from hadron and muon components is heavier than that which can be 
obtained using mainly electromagnetic and muon components. Muons are usually less 
model dependent at the fixed primary energy, since they are penetrating particles and 
are collected from all atmospheric altitudes representing something like an integral 
over the longitudinal profile of the shower. Taking them as the basis for the 
comparison we should expect that for well tuned, consistent models the analysis of 
ratios $N_e/N_\mu$ and $N_h/N_\mu$ should give the same $\langle lnA \rangle$.  The 
larger  $\langle lnA \rangle$ value (2.25$\pm$0.08) found in KASCADE experiment from 
$N_h/N_\mu$ analysis than that from $N_e/N_\mu$ (1.90$\pm$0.05) \cite{Roth} indicates
 that the difference between $N_e/N_\mu$ and $N_h/N_\mu$ in the present models is too 
low (~Figure 1~). In the more realistic model this difference should be increased.
\begin{figure}[htb]
\begin{center}
\includegraphics[width=7.5cm,height=7.5cm]{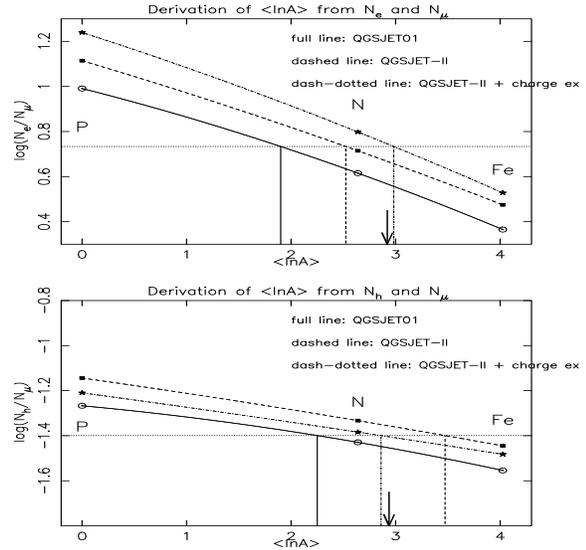}
\caption{\footnotesize The illustration of the way to get the consistent value of mean 
logarithmic 
mass $\langle lnA \rangle$ from (a) $N_e/N_\mu$ and (b) $N_h/N_\mu$ measurements. 
Initial values of $N_e/N_\mu$ and $N_h/N_\mu$ are calculated for 1 PeV primary protons,
nitrogen and iron nuclei using the QGSJET01 model and shown by the full line. 
The dashed line shows the same ratios for the QGSJET-II model: $N_e/N_\mu$ and 
$N_h/N_\mu$ rise by the same 
factor. The dash-dotted line demonstates the effect of the charge exchange combined 
with the QGSJET-II: $N_e/N_\mu$ rises, but $N_h/N_\mu$ falls. The thick arrow indicates
the consistent value of $\langle lnA \rangle$.}
\end{center}
\label{fig:eas1}
\end{figure}
 
The theoretical basis of QGSJET-II model besides the reduction of the interaction cross
 section for pions and an inelasticity requires also the reduction of the multiplicity 
of the secondary particles. Simulations of EAS showed that transition from QGSJET01 
to QGSJET-II model at the fixed energy leads to the {\bf rise} of $N_e/N_\mu$ and 
$N_h/N_\mu$ ratios both for primary protons and for all primary nuclei. 
It is because the EAS electromagnetic and hadron components follow each other in the 
lower part of the atmosphere, i.e. beyond the shower maximum, in an approximate 
equilibrium. However, the difference between $\langle lnA \rangle$ values derived from
$N_e/N_\mu$ and $N_h/N_\mu$ ratios in the QGSJET-II model becomes even larger than for 
QGSJET01. While the $\langle lnA \rangle$ value derived from $N_e/N_\mu$ ratio rises 
from 1.9 to 2.52, that derived from $N_h/N_\mu$ rises from 2.25 to 3.47 (~see Figure 
1~). Therefore, the reduction of the interaction cross-section for pions, of the 
inelasticity and the multiplicity in the QGSJET-II, does not remove the existing 
difference between $\langle lnA \rangle$ values and the inconsistency still holds. 

To eliminate this inconsistency we have suggested the additional
transfer of the energy into an electromagnetic component in the cascading 
process \cite{EW2}. It has been made on a pure phenomenological
basis. Later we have suggested the so called 'sling 
effect' in nucleus-nucleus interactions as the process responsible for a deeper 
penetration into the atmosphere of cascades induced by primary nuclei and also for 
an additional electromagnetic component in them \cite{EW3,EW4}. This effect could serve
 as a possible theoretical basis of the needed model modifications. However, this 
effect being very probable, seems to be small to give the noticable 
changes. The needed effect has to be stronger and we now think that the charge exchange
 process is the likely culprit.

In fact charge exchange processes are already taken into account in both QGSJET01 and 
QGSJET-II models. The question is whether it is possible to modify the probability of 
this process without a conflict with the existing experimental data and whether such a 
modification could give the consistent $\langle lnA \rangle$ value both from 
$N_e/N_\mu$ and $N_h/N_\mu$ ratios. 

In order to analyse this possibility I used the option provided by the HDPM model. It 
is the only model within the CORSIKA code where one can switch on and off the charge 
exchange processes and by this way to estimate the effect which this process has on 
$N_e/N_\mu$ and $N_h/N_\mu$ ratios. Actually I determined the ratios 
$R_{e\mu} = (N_e/N_\mu)_{HDPM_+}/(N_e/N_\mu)_{HDPM_-}$ and
$R_{h\mu} = (N_h/N_\mu)_{HDPM_+}/(N_h/N_\mu)_{HDPM_-}$ for the HDPM model with and
without the charge exchange process (~denoted as $HDPM_+$ and $HDPM_-$ respectively~).
Then I applied these ratios to the QGSJET-II model as 
$log(N_{e,h}/N_\mu)_{QGSJET-II+ch.exch.}=log(N_{e,h}/N_\mu)_{QGSJET-II}+\varepsilon logR_{e,h}$. 
Here the coefficient $\varepsilon$ limited by $0 \leq \varepsilon \leq 1$ 
gives an estimate of possible increase of the charge exchange needed to get a 
consistent value of $\langle lnA \rangle$ derived from $N_e/N_\mu$ and $N_h/N_\mu$ 
ratios, denoted as $\langle lnA \rangle_{e\mu}$ and $\langle lnA \rangle_{h\mu}$ 
respectively.          

This exersise shows that by this way it is possible to achieve the needed consistency.
Simulations show that inspite of the addtionsl energy transfer the shower size at the 
sea level remains practically the 
same as in the absence of the charge exchange process, whereas the number of hadrons 
and muons decreased. So that in contrast to the modifications provided only by 
non-linear effects in the QGSJET-II model when both $N_e/N_\mu$ and $N_h/N_\mu$ rise, 
the increase of the charge exchange probability leads to the rise of $N_e/N_\mu$, but 
to the fall of $N_h/N_\mu$, because both $N_\mu$ and $N_h$ fall but the latter falls 
stronger. For example, with  $\varepsilon = 1$ $\langle lnA \rangle_{e\mu}$ rises from 
2.52 to 2.98, but $\langle lnA \rangle_{h\mu}$ falls from 3.47 down to 2.86 (~see 
Figure 1~). This 'overshooting' is due to that the charge 
exchange process is already taken into account in the QGSJET-II model and 
application of the expression given in the previous paragraph with $\varepsilon = 1$ 
makes this process too strong. The consistent value of  
$\langle lnA \rangle_{e\mu} = \langle lnA \rangle_{e\mu} = 2.94 \pm 0.09$ is achieved 
at $\varepsilon = 0.88 \pm 0.12$. The errors are derived from the statistical errors of
mean values of $N_e/N_\mu$ and $N_h/N_\mu$ ratios obtained by Monte-Carlo simulations 
with CORSIKA6.014 and CORSIKA6500 codes and taking the values of 
$\langle lnA \rangle_{e\mu}$ = 1.90$\pm$0.05 and $\langle lnA \rangle_{h\mu}$ = 
2.25$\pm$0.08 obtained from the experimental data using the QGSJET01 model \cite{Roth}.
 Systematic errors are difficult to evaluate at
this stage of analysis, but actually our estimates are given just to demonstrate the 
principal opportunity to use the QGSJET-II model with an enhanced charge exchange 
probability to get a consistent estimates of the primary CR mass composition. 

The higher consistent value of $\langle lnA \rangle$ = 2.94$\pm$0.09 compared with 
values of 1.90$\pm$0.05 and 2.25$\pm$0.08 derived from $N_e/N_\mu$ and $N_h/N_\mu$ 
ratios with the QGSJET01 model means that the true primary CR mass composition should 
be heavier in the new analysis.
    
Simulations show also that with the increased charge exchange the depth of maximum for
 proton induced showers shifts upwards by about 5 gcm$^{-2}$, but for nitrogen and iron
 induced showers it moves downwards by 1 gcm$^{-2}$ and 5 gcm$^{-2}$ respectively. 
Hence the increase of the charge exchange cannot destroy the deeper penetration of
EAS by about 20 gcm$^{-2}$ provided by the QGSJET-II model. 
    
\subsection{The inelastic charge exchange for pions}

Now we shall discuss the process which could be used to increase the charge exchange
probability. At the end of sixties one of
 the inventors of the ionization calorimeter, V.S.Murzin from the Moscow 
University used this detector to study interactions of CR pions. He has found
 that with a considerable probability the charged pion could lose its charge but 
preserve a good part of its initial energy. He called this process 'the inelastic 
charge exchange' \cite{Murzi}. Actual numbers were the following: the probability to 
preserve more 
than 0.5 of the energy in the collision of the pion with iron nuclei has been estimated
 as 10\% and this probability seemed to be independent of energy in hundred GeV - TeV 
energy range. 

Since then numerous experiments have been made on the production of neutral pions in 
hadron interactions. Their results can be found in the Particle Data Group archive 
(~http://durpdg.dur.ac.uk/HEPDATA/reac.html~) (~see also the list of literature in 
\cite{Ataya}~). However, measurements of collisions with nuclei are still 
sparse and cover mostly the high transverse momentum (~$P_t$~) region. The relevant 
data relate mostly to $\pi^{\pm} P$ interactions and often give the values of invariant
 cross-sections \cite{Ataya,Aguil,Pauss,Abdur,Apsim,Barne}. After integrating over
 $P_t$ some extreme examples \cite{Aguil,Pauss,Apsim} of the obtained inclusive spectra
 for neutral pions are shown in Figure 2. 
\begin{figure}[htb]
\begin{center}
\includegraphics[width=7.5cm,height=7.5cm,angle=0]{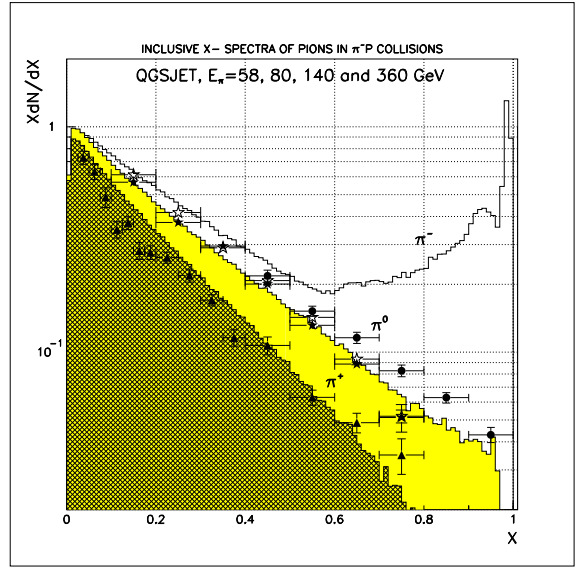}
\caption{\footnotesize The inclusive spectrum of neutral pions produced in $\pi^- P$ 
interactions. Full circles - for the 58 GeV $\pi^-$ beam energy \cite{Pauss}, open and 
full stars -  for 80 and 140 GeV respectively \cite{Apsim}, triangles - for 360 GeV
\cite{Aguil}.  Horizontal error bars indicate the interval of 
$X$ used for the determination of the invarint cross-section. Histograms are spectra of
 $\pi^+, \pi^0$ and $\pi^-$ calculated with CORSIKA INTTEST version for the 
QGSJET01+GHEISHA interaction model at 80 GeV. The use of the QGSJET01 model at this 
energy is justified since non-linear effects introduced in QGSJET-II appear at much 
higher energies.}
\end{center}
\label{fig:eas2}
\end{figure}
   
It is seen that the results of different experiments have a considerable spread. 
While the result of one experiment matches the model perfectly \cite{Apsim}, the 
 $\pi^0$-spectrum obtained in another experiment follows the spectrum of $\pi^+$
\cite{Aguil} and 
there is a spectrum which is definitely higher than that which matches the model 
\cite{Pauss} although they agree with each other within error bars. This 
difference cannot be caused by the energy dependence of the process since models and 
experiments show no appreciable dependence in the 58 - 360 GeV energy interval 
\cite{Barne}. The examination of these data indicates that the probability to have 
$\pi^0$ with $X > 0.5$ is not 10\%, but about 7\% at these energies and for $X > 0.7$ 
it is about 2.3\%.      

 It seems that on the basis of the spread of experimental data and the 
absence of the preference between the different results the model can be
re-tuned according to indications of the EAS analysis towards the higher probability of
 the charge exchange process for pions. This can improve the consistency between 
$\langle lnA \rangle$ values obtained from different EAS components. Since at $X > 0.7$
the difference between different experiments rises up to $\sim 2.5$ and continues 
growing with $X$, the big value for the estimate of $\varepsilon = 0.88 \pm 0.12$, 
presented in the previous subsection, is not inconsistent with these experimental data.
       
There is another point which should be remarked. The appearance of leading neutral 
pions in charged pion collisions observed in cosmic-ray and in some of accelerator 
experiments is interpreted within the framework of the triple-region description with a
substantial contribution of RRP-term. Its contribution should give the flat behaviour 
of the inclusive $\pi^0$ cross-section at large $X > 0.7$. It is not seen in the data 
shown in Figure 2. This discrepancy is not clear. However, in can be that $\pi^0$'s 
with very large $X$ are biased in accelerator experiments by triggering conditions, 
with partially suppressed low multiplicity and diffraction events as has been mentioned
 in \cite{Aguil}, while the introduced corrections are model dependent. There was no 
such bias in cosmic-ray experiments.
  
The experimental value of the mean fraction $\alpha_\gamma$ of energy transferred by 
$\pi^-$ to $\pi^0$ is independent of energy and equal to 0.25$\pm$0.01 \cite{Murzi}.
The $\alpha_\gamma$ provided by the QGSJET01 model shows a slight decrease with the 
energy and above 10 TeV falls below the experimental value (~Figure 3~). Independently 
of the origin of this fall (~e.g. change of the mass composition of secondaries~) it is
 another indication that the energy transfer into the electromagnetic component should 
be slightly increased to improve the model. 
\begin{figure}[htb]
\begin{center}
\includegraphics[width=7.5cm,height=7.5cm,angle=0]{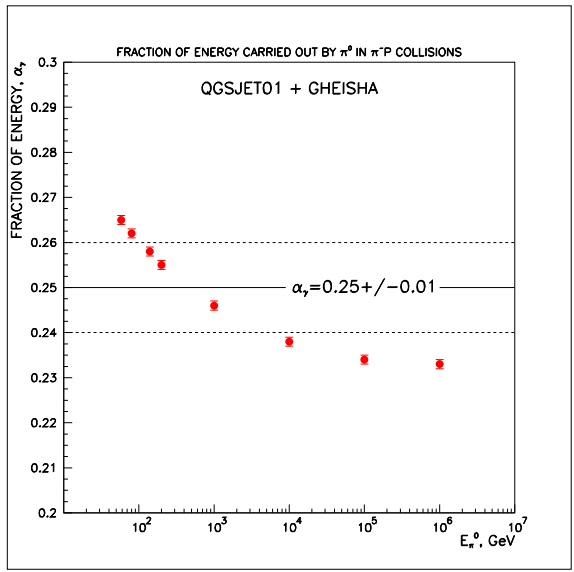}
\caption{\footnotesize The mean fraction of energy $\alpha_\gamma$ transferred to neutral pions in $\pi^-P$ 
collisions as a function of energy. The experimental value of $\alpha_\gamma =0.25\pm0.01$
is from \cite{Murzi} and denoted by full and dashed lines. The full circles show
$\alpha_\gamma$ values calculated with the QGSJET01+GHEISHA model. The calculated values are 
slightly smaller than experimental ones above 10 TeV.}  
\end{center}
\label{fig:eas3}
\end{figure}

\section{Interactions of EAS with detectors and an environment}

The second part of this paper relates to interactions of EAS particles not with air 
nuclei, but with another target: the stuff of the detectors, their environment and the 
ground. The aim of this part is to draw an attention to the possible contribution of 
low energy neutrons created in such interactions to the signal in the hydrogen 
containing detectors, deployed particularly at mountains. 

It is well known that the thickness of the shower disk depends on the distance 
from the shower core rising from 2-3 m at the EAS center to tens of m at about 1 km 
from it. It corresponds to the time 'thickness' of a few hundreds ns. The delayed 
particles which appear a few microseconds after the main shower front were observed and
 discussed long ago \cite{Tongi,Greis,Linsl}. However, the discovery of neutrons 
delayed by {\em hundreds} microseconds in the shower core made by Chubenko A.P. and his
 group with the Tien-Shan neutron monitor \cite{Chub1,Aushe,Anto1,Anto2,Chub2} seemed 
to be unexpected and attracted an attention \cite{Sten1,Sten2,Gawin,Baygu,Jedrz}. 

Though the dispute on the interpretation of this finding is not finished the majority 
of participants is inclined to explain it by the interaction of hadrons in the EAS core
 with the stuff of the detector, i.e. by multiplicative processes in the lead of the 
neutron monitor with the subsequent long diffusion and thermalization of released 
neutrons in the monitor's moderator and reflector \cite{Jedrz,Sten3,Erly4} (~see, 
however, another view in \cite{Chub3}~). Independently of the interpretation the 
phenomenon is very spectacular and looks as the neutron 'thunder' which appears with 
a time delay after the 'lightning' which is the EAS itself \cite{Erly4,Erly5}. 

If the interpretation of the majority is correct, the observed neutrons can be produced
also in the ground which is not so heavy as lead, but nevertheless there are many 
heavy elements in it (~mainly Si~) which could produce neutrons being hit by an EAS core. 
There is also water in it which serves as a good moderator like in nuclear reactors.
 The influence of the ground and ground-based environment has to be more substantial 
at the mountain level, where EAS cores are much more energetic and a good part of the year 
the ground is covered by snow. Sometimes this snow is of meters thick (~Tien-Shan, Aragats, 
Chacaltaya, South Pole etc.~). As for the Tien-Shan station an additional factor is 
that it is built on the permafrost with a good part of ice in it. Since neutrons can 
diffuse up to long distances from the place where they are produced \cite{Agafo} their 
effect might be noticable even at shallow depths underground. 

Many running EAS arrays use water or ice cherenkov detectors: Pierre Auger 
Observatory, Milagro, Nevod, Ice-Top etc.  At the first sight they should not be 
sensitive to neutrons, since they are neutral particles and mostly non-relativistic. 
However, the experimental study of the neutron 'thunder' revealed that delayed neutrons
 are accompanied by gamma-quanta and electrons \cite{Anto2}, which in principle could 
give a signal in water tanks.  

It is particularly relevant to Pierre Auger Observatory. The comprehensive modelling of
 the effect of albedo neutrons emitted by the ground as a result of the EAS interaction
 is complicated and planned for the future paper. Here I show that the effect can be 
noticable even taking into account only EAS neutrons. 

Simulations of the EeV proton
 induced showers observed at the altitude of 1400 m a.s.l. show that neutrons are the 
most abundant among EAS hadrons and their lateral distribution function (~LDF~) is 
wider than LDF for protons and pions. At the typical distance of $\sim$1 km from the 
core the density of neutrons with energy above 50 MeV and their energy density is about
 the same as for muons of this energy and about 10\% of the gamma-quanta plus electrons
 with energy above 1 MeV (~Figure 4~).        
\begin{figure}[htb]
\begin{center}
\includegraphics[width=7.5cm,height=7.5cm,angle=0]{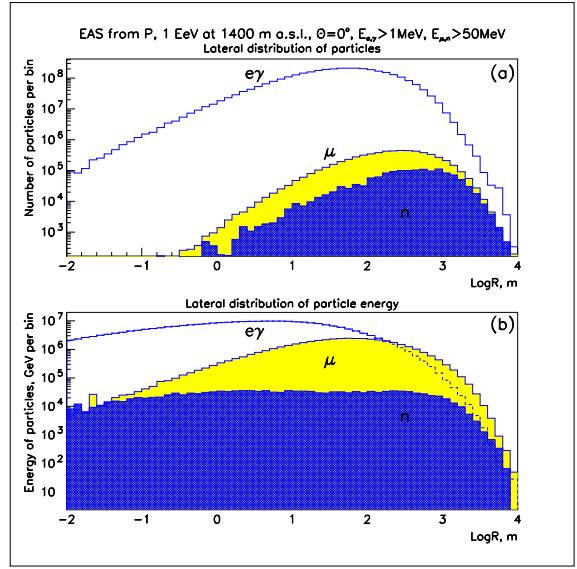}
\caption{\footnotesize Lateral distribution of particle numbers (a) and the particles 
energy (b) for
1 EeV primary proton incident vertically at the level of 1400 m a.s.l. It is seen that 
neutrons could contribute up to 10\% to the signal of water tanks at 1km distance from 
the shower axis, if among products of their interaction with water are relativistic 
electrons.} 
\end{center}
\label{fig:eas4}
\end{figure}

Moreover after about 5 $\mu s$ behind the EAS front the neutron component at 1 km from 
the EAS axis becomes dominant, overtaking muons, electrons and gamma-quanta (~Figure 
5~).  
\begin{figure}[htb]
\begin{center}
\includegraphics[width=7.5cm,height=7.5cm,angle=0]{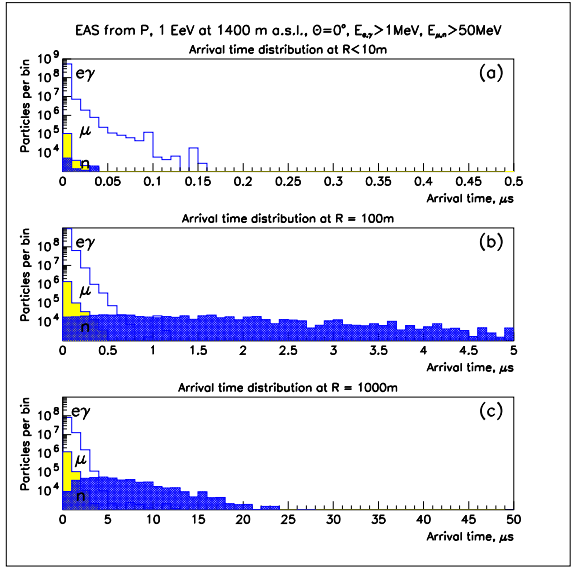}
\caption{\footnotesize Arrival time distribution of electromagnetic, muon and neutron component of 
the shower at core distances less than 10m (a), 100m (b) and 1000m (c). It is seen that
 at 1000m from the core neutrons dominate among other particles after 5 $\mu s$.} 
\end{center}
\label{fig:eas5}
\end{figure}

This distance and the time delay are right the working distances and times for 
Pierre Auger Observatory, so that the possible contribution of neutrons to the signal 
from their water tanks should be analysed and taken into account if necessary. The same
 remarks could be referred to hydrogen containing plastic scintillators used in many 
other large arrays (~Yakutsk, Telescope Array etc.~). As it has been mentioned above, 
signals delayed by $\mu s$ (~'subluminal pulses'~) have been already observed in large 
scintillator arrays, such as Volcano Ranch \cite{Greis,Linsl}. 

A good analysis of the 
possible effect of delayed particles on the primary energy estimation has been made in
\cite{Dresc} applicable to the AGASA array. It has been shown that overestimate of the 
primary energy for its scintillators and the acquisition system cannot exceed a few 
percents. For other arrays it may be higher.   

Presumably the effect of 'the neutron thunder' can be applied in practice for 
the neutron carotage of the upper layers of the ground. Instead of the artificial 
neutron source in this method the ordinary EAS can be used since EAS cores carry on 
and produce a lot of secondary neutrons. Also 'the neutron thunder' can be used for 
the search of water on the Moon \cite{Feld} or on the surface of other
planets \cite{Mitr1,Mitr2}.       
  
\section{Conclusion}

The analysis of existing controversies in the interpretation of
experimental data on EAS indicates that an improvement of our
understanding of the EAS phenomenon and the self-consistency of results
on primary CR derived from EAS can be achieved by a moderate
modification of the current particle interaction models. This
modification has to result in a slightly deeper penetration of EAS
into the atmosphere as well as in the increased transfer of the
energy from the hadronic to electromagnetic components of EAS. The
account for non-linear effects in parton interactions like that in
the QGSJET-II model and an increasing probability of inelastic charge
exchange processes for pions can help.

Here it is appropriate to make some general remarks.
The nuclear and electromagnetic nature of EAS has been established at
 the end of forties. That was the time when the world greatest
 accelerator - Dubna Synchrophasotron had not been commissioned and CR 
 were the unique source of information about high energy interactions. It is 
surprising that now, after about 50 years of the leading role of accelerators in the 
field, CR are still able to contribute to our understanding of the high energy
 interactions. Another point is that after nearly 70 years since the discovery of EAS
by Pierre Auger and Roland Maze we still develop our understanding of this phenomenon.
The discovery of the 'neutron thunder' certainly complements our knowledge of the EAS
development and is worth of further experimental and theoretical study.
    
{\bf Acknowledgments}

The author thanks the INFN, sez. di Napoli and di Catania, personally Professors 
M.Ambrosio and A.Insolia for providing the financial support for this work and their 
hospitality. I also thank Capdevielle J.N., Martirosov R., Ostapchenko S., Petrukhin 
A., Ryazhskaya O.G., Stanev T., Szabelski J., Ter-Antonian S., Tsarev
V.A., Watson A. and Yodh G. for useful discussions and references.

\end{document}